\RequirePackage{fix-cm}
\documentclass[twocolumn]{svjour3}          % twocolumn
\smartqed  % flush right qed marks, e.g. at end of proof
\usepackage{graphicx}
\journalname{Computing and Software for Big Science}
\begin{document}

\title{HEPCloud, a new paradigm for HEP facilities: CMS Amazon Web Services Investigation}

\author{Burt~Holzman\thanks{Corresponding author: burt@fnal.gov} \and Lothar~A.T.~Bauerdick \and Brian~Bockelman \and Dave~Dykstra \and Ian~Fisk \and Stuart~Fuess \and Gabriele~Garzoglio \and Maria~Girone \and Oliver~Gutsche \and Dirk~Hufnagel \and Hyunwoo~Kim \and Robert~Kennedy \and Nicolo~Magini \and David~Mason \and Panagiotis~Spentzouris \and Anthony~Tiradani \and Steve~Timm \and Eric~Vaandering}

\institute{B.~Holzman \and L.A.T.~Bauerdick \and D.~Dykstra \and S.~Fuess \and G.~Garzoglio \and O.~Gutsche \and D.~Hufnagel \and H.~Kim \and R.~Kennedy \and N.~Magini \and D.~Mason \and P.~Spentzouris \and A.~Tiradani \and S.~Timm \and E.~Vaandering \at
              Fermi National Accelerator Laboratory
           \and
           B.~Bockelman \at
              University of Nebraska, Lincoln
           \and
           I.~Fisk \at
              Simons Foundation
           \and
           M.~Girone \at
              CERN
}

\date{Received: date / Accepted: date}

\maketitle

\begin{abstract}
Historically, high energy physics computing has been performed on large purpose-built computing systems. These began as single-site compute facilities, but
have evolved into the distributed computing grids used today. Recently, there has been an exponential increase in the capacity and capability of
commercial clouds. Cloud resources are highly virtualized and intended to be able to be flexibly deployed for a variety of computing tasks.
There is a growing interest among the cloud providers to demonstrate the capability to perform large-scale scientific computing.
In this paper, we discuss results from the CMS experiment using the Fermilab HEPCloud facility,
which utilized both local Fermilab resources and virtual machines in the Amazon Web Services Elastic Compute Cloud.
We discuss the planning, technical challenges, and lessons learned involved
in performing physics workflows on a large-scale set of virtualized resources. In addition, we will discuss the economics and operational efficiencies
when executing workflows both in the cloud and on dedicated resources.


\end{abstract}

\section{\label{sec:introduction}Overview}

The use of highly distributed systems for high-throughput computing has been very successful for the broad scientific computing
community.  Programs such as the Open Science Grid \cite{osg} allow scientists to gain efficiency by utilizing available cycles across different domains. Traditionally, these programs have aggregated resources owned at different institutes, adding
the important functionality to elastically contract and expand resources to match instantaneous demand as desired.
An appealing scenario is to extend the reach
of extensible resources to the rental market of commercial clouds.

A prototypical example of such a scientific domain is the field
of High Energy Physics (HEP), which is strongly dependent on high-throughput computing.
Every stage of a modern HEP experiment requires massive resources (compute, storage, networking). Detector and simulation-generated data have to be processed and associated with auxiliary detector and beam information to generate physics objects, which are then stored and made available to the experimenters for analysis. In the current computing paradigm, the facilities that provide the necessary resources utilize distributed high-throughput computing, with global workflow, scheduling, and data management, enabled by high-performance networks. The computing resources in these facilities are either owned by an experiment and operated by laboratories and university partners
(e.g.\ Energy Frontier experiments at the Large Hadron Collider (LHC) such as CMS, ATLAS) or deployed for a specific program, owned and operated by the host laboratory
(e.g.\ Intensity Frontier experiments at Fermilab such as NOvA, MicroBooNE).

\begin{figure}
\centering
\includegraphics[width=.5\textwidth]{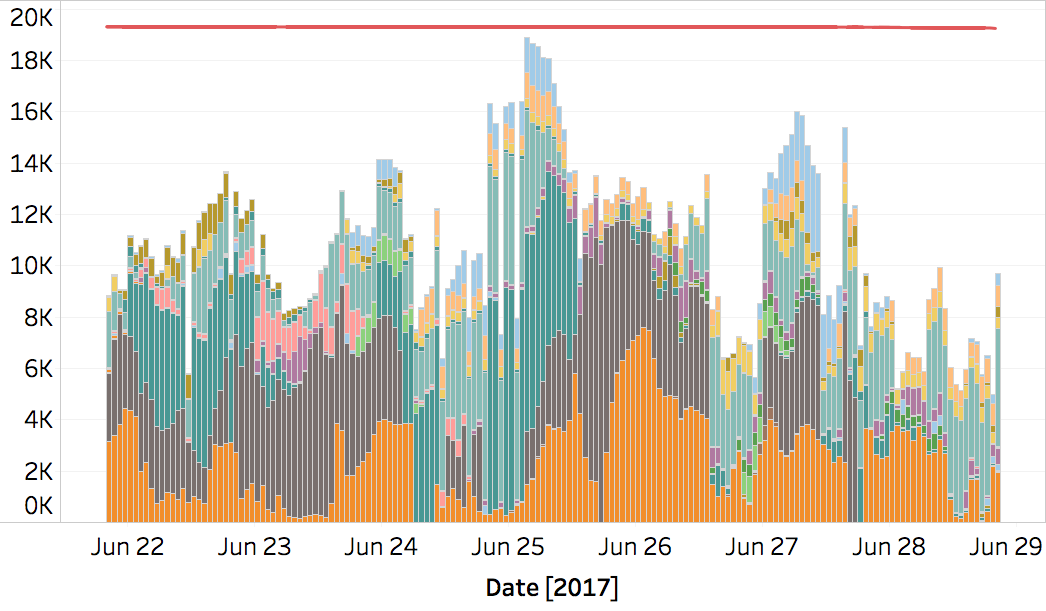}
\caption{\label{fig:burst}Utilization of the Fermilab general purpose scientific computing cluster. The different colored bars correspond to different
experiments; the line represents the total number of available cores.}
\end{figure}

The HEP investment to deploy and operate these resources is significant: for example, at the time of this work, the size of the worldwide computing infrastructure for the CMS experiment at the LHC is 150,000 cores, with US CMS deploying 15,000 cores at the Fermilab Tier-1 site, and more than 45,000 cores at seven Tier-2 sites. 
Computing activity of a HEP experiment like CMS can be separated into organized processing activities that are planned and centrally managed,
and analysis activities that are chaotic, submitted by many individuals and much less predictable. Common to both, all computing activity
comes in ``bursts'', depending on the accelerator schedule and the availability of new advances in software and understanding of the
detector (calibration and alignment). A typical week for an on-premises Fermilab cluster across multiple experiments is shown in Figure \ref{fig:burst}.
This demonstrates the stochastic nature of compute demand
as well as the expense of provisioning for peak capacity rather than steady-state use. At the same time, peak capacity is required to do large amounts of computing
in a short amount of time (e.g. new analyses with discovery potential, conference deadlines).

The evolution of the HEP experimental program (upgrades, new experiments) will generate increased computing needs that go well beyond any performance gains expected from  advancements in computing techniques and technologies. Furthermore, due to power and cooling requirements, new architectures are departing from Moore's law expectations, resulting in lower performance per core. By 2025, the muon program, the long-baseline and short-baseline neutrino programs, and the LHC will be at the apex of their offline analysis, as the two new programs (High-Luminosity LHC, DUNE) are coming on-line producing massive amounts of data \cite{lhc,hl_lhc,p5}. The increased precision, event complexity, and luminosity of the HL-LHC alone will push computing needs nearly two orders of magnitude above current HEP capabilities, while generating exabytes of data.  The implied increase in compute capacity may not be possible to satisfy with on-premises homogeneous
resources.
As a result of this as well as the stochastic demand, we need infrastructure to manage heterogeneous distributed resources, optimize their use, and maximize their efficiency and cost effectiveness.

It is essential for HEP to develop the concepts and deploy the infrastructure that will enable analysis of these vast amounts of data efficiently and cost effectively.
Following the Infrastructure-as-a-Service (IaaS) paradigm \cite{Cloud}, US HEP facilities could incorporate and manage rental resources, achieving elasticity that satisfies demand peaks without over-provisioning local resources.

Along this paradigm, the HEPCloud facility concept is envisioned to be a portal to an ecosystem of heterogeneous commercial and academic computing resources.
It will provide ``complete solutions'' to users with agreed-upon levels of service, automatically routing user workflows to on-premises and off-premises resources
based on efficiency metrics, cost, workflow requirements, and the policies of the facilities hosting the resources. This will be done transparently to the user, utilizing a sophisticated decision engine and cost model, and policies for managing user allocations to potential computing resources. This includes managing
security and access controls to leadership-class computing facilities on behalf
of the user community.

In order to investigate the merit of this approach, we deployed the Fermilab HEPCloud pilot project. The goal was to produce a design of the overall concept and deploy a first
implementation of important components, allowing us to investigate the merit of this approach, and to evaluate different solutions. 
The objective was to integrate rented resources into the current Fermilab computing facility in a manner transparent to the user.  The first type of external resources considered was commercial clouds, through partnerships with different providers, and the first partnership was with Amazon Web Services (AWS). For our studies, we identified use cases that both emphasized and exercised the necessary aspects of the concept and were also useful to the experimenters. One of these use cases focused on CMS Monte Carlo generation and reconstruction, targeting physics results for the Moriond conference in March 2016 \cite{MoriondQCD16,MoriondEW16,cmsresults}. This use case studied the scalability and sustainability of elastic provisioning of AWS resources through the portal, and exercised the prototype decision engine and cost model.

The HEPCloud concept is a facility-centric approach to computing resources; a host facility (Fermilab) manages the resource acquisition decisions.  The HEP field has made significant inroads in utilizing elastic resources.  Earlier work includes \cite{atlas_cloud,atlas_cloud2,belle_cloud,cms_cloud}; contrasting with the HEPCloud approach, these papers survey the status of cloud use and cost comparisons from an experiment point-of-view.

\section{CMS Use Case: Introduction}

The CMS experiment is facing a large and ever increasing computing challenge. To meet the growing computing needs, CMS has considered the use of resources beyond the traditional CMS owned grid-provided systems.

One appealing potential solution is the utilization of dynamically provisioned resources---either via academic and opportunistic access, or through commercially provided computing services.  The logical
platform choice for a first implementation of this solution was the market leader \cite{Gartner_2015} in cloud IaaS, Amazon Web Services (AWS).
CMS demonstrated small-scale cloud computing on this platform for a short amount of time as a proof-of-concept to investigate feasibility \cite{gwms_cloud,evans11}.
%Beginning in 2015, CMS began to seriously explore commercial cloud provisioned resources.
The HEPCloud demonstration described herein
took the next big step and was intended to show the ability to increase the global processing capacity of CMS by a significant fraction---60,000 cores---for an extended period. Importantly, the test was also intended to deliver useful simulated physics events to the collaboration for analysis at a production scale.

To deploy resources for the use case, CMS was awarded a 9 to 1 matching grant from AWS that allowed the purchase of \$300k of credits for computing, storage, and network charges for an investment of \$30k.
The size of the award was based on an estimate of what it would cost to do one month of large-scale processing.
Additionally, a conditional cost waiver was granted for exporting data; as long as the export costs remained under 15\% of the total monthly bill, and were transmitted across research networks such as ESNet, the export charges would be waived entirely. This discount program was so successful that it has been extended to researchers
at all academic and research institutions \cite{CostWaiver}.

CMS has three standard workflow types that it has traditionally executed on dedicated resources.
{\it GEN-SIM} generates physics events via pseudo-random number generators, with no input files and large output files;
{\it DIGI-RECO} simulates the detector response and reconstructs physics quantities (e.g. tracks with particle identification), with large input files and modest output files; and
{\it DATA-RECO} reconstructs physics quantities from detector data, with large input files and modest output files.
We chained together {\it GEN-SIM} and {\it DIGI-RECO} into a {\it GEN-SIM-DIGI-RECO} workflow, an optimal case for minimizing egress charges.
Over the last six months, {\it GEN-SIM} and {\it DIGI-RECO} represented more than half of the global CPU resources utilized by the experiment.
We considered many different physics workflows to simulate for the experiment, evaluated the applicability of each solution, and the urgency of the scientific needs. We chose four {\it GEN-SIM-DIGI-RECO}
workflows (``TTJets'', ``DY\_M10--50'', ``DY\_M50'', ``WJetsToLNu'') that were both needed and judged to be most appropriate for the test.

In the following sections, we discuss the tests performed, the services required, and the scale and performance achieved. We also evaluate the cost to provide dedicated computing resources at Fermilab versus the costs paid to AWS for the same capacity.

\section{Procurement Evaluation}
The Fermilab HEPCloud team, in consultation with CMS staff and the Fermilab Procurement office, wrote a set of specifications for commercial cloud providers. This included a set of financial and technical requirements to satisfy the need of the Fermilab HEPCloud facility project. The financial requirements included the ability to track spending by groups, to account for regular spending and credits, to access technical support, and to pay in advance with pro forma invoicing. The technical requirements included access to a minimum scale of storage and CPU cores, guaranteed network bandwidth to the ESNet science network, support for certain APIs to launch virtual machines, support for monitoring, and the ability to alert on preset levels of spending.  A Statement of Work was prepared that described the activity of CMS and three other projects, and specified a total amount of computing services that would be purchased.  A wide variety of cloud providers and resellers were requested to bid on this Request For Quotations (RFQ) \cite{FNAL_RFP}. 
The bids were evaluated according to a predetermined set of best value criteria. Once the qualifying bid was selected, the team certified that the services of the selected provider met the requirements of the RFQ. DLT Solutions, a reseller of Amazon, was awarded the contract with terms that give us the flexibility to immediately spend funds when they are available.

Major cloud vendors offer grants for customers to explore their platform. For this work, Fermilab has obtained two Amazon Web Services Research Grants for the NOvA and CMS experiment.
While general discussions on the users' needs and the grant programs are conducted freely with the vendors, the acceptance of the grant and related service access need to be coordinated with procurement.
This ensures fairness in the competitive bidding process and fosters compliance with government restrictions on undersigning customer agreements.
As for the latter, resellers of cloud services provide government-friendly contracts by taking upon themselves the liability of looser indemnification clauses.
Purchased services and grant credits can be managed together as a single program.

\section{Services}
\subsection{Services Deployed in AWS}

\begin{figure}
\centering
 \includegraphics[width=.5\textwidth]{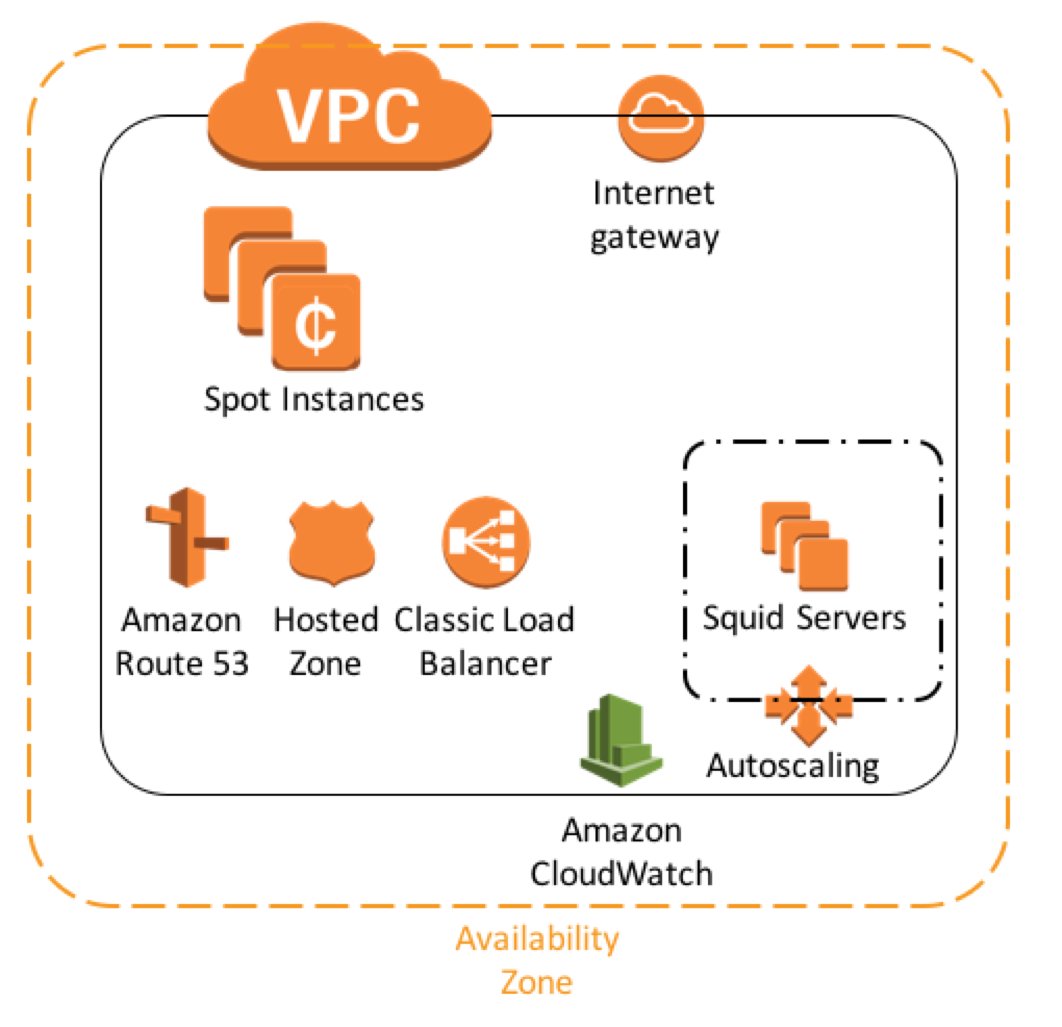}
\caption{\label{fig:aws_arch}A simplified overview of the AWS services utilized for the CMS use case.}
\end{figure}

A simplified overview of the architecture of the AWS services used is shown in Figure \ref{fig:aws_arch}. These components are described in detail in the
following sections.

\subsubsection{Caching Application Code and Database Access}
The sole on-demand service deployed for the CMS use case was a dynamically scaled web cache.  This service ran Squid servers\footnote{Optimizing Web Delivery, http://www.squid-cache.org} on the front end, utilizing CERN-VMFS (CVMFS)
technology \cite{cvmfs} to cache the application code, and the Frontier service \cite{frontier} to cache database queries from an off-site experimental conditions
database. From previous studies \cite{Timm_chep_2015}, it was observed that a squid service that was not co-located with compute nodes had a latency that was high enough to degrade performance. In order to instantiate the service, AWS CloudFormation\footnote{https://aws.amazon.com/cloudformation/} was used to orchestrate the entire process for launching and tearing down the service and dependent infrastructure. The following services and/or components were configured by the the CloudFormation template created by the Fermilab HEPCloud team:
\begin{itemize}
\item Elastic Load Balancer (ELB)\footnote{https://aws.amazon.com/elasticloadbalancing/}
\item Auto Scaling Group\footnote{https://aws.amazon.com/autoscaling/} (from 1 to N servers) and policies
\item Squid-optimized Amazon Machine Image
\item Route 53\footnote{https://aws.amazon.com/route53/} DNS CNAME Record Set
\item CloudWatch\footnote{https://aws.amazon.com/cloudwatch/} alarms
\end{itemize}

One CloudFormation stack as described above was deployed per availability zone. Each stack had a known, fixed address, which was predetermined by the Route 53 service. An initialization script ran in each virtual machine at launch that auto-detected which region and availability zone it was in, as well as the public hostname of the machine. CMS-specific configuration files
were modified at launch to point to the appropriate squid server and the location of the input data via an AWS Simple Storage Service (S3)\footnote{https://aws.amazon.com/s3/} URI.

The squid server was used to cache both the experiment-specific software, which was delivered via CVMFS, and event-specific database information, which came from Frontier. CVMFS maintains its own local cache on the local disk, while Frontier does not. In order to efficiently cache the Frontier information, we instantiated a squid cache on each worker node, reducing the net load on the AWS central squid servers and on the main Frontier servers at CERN, as well as the cost of traffic through the Elastic Load Balancer (ELB).
At peak load, each ELB stack was running eight squid servers, with four squid processes per server. Each server was instantiated on a 4-core c3.xlarge instance \cite{aws_types}.
Eight stacks were instantiated, each one corresponding to a different region/zone combination.

\subsubsection{AWS Network Configuration}

Each AWS region has a default Virtual Private Cloud (VPC) that defines the network configuration for an AWS account. The default VPC was modified to accommodate the CMS use case. The VPC is set up with one Internet Gateway configured with a subnet per Availability Zone. Once the VPC is configured, Security Groups act as firewalls to control the traffic allowed to reach the virtual machines. Two Security Groups were configured. One Security Group allowed the squid servers to contact the general internet in order to retrieve data and cache it. The other Security Group restricted outbound network access to Fermilab, CERN, and the AWS public IP addresses; this group allows inbound ssh access from Fermilab only.

\subsubsection{AWS Spot Instances}

AWS sells their excess resource capacity following a market model called ``Spot Market''. For every combination of machine type, availability zone, and region, users supply a bid price that represents the maximum that they are willing to pay per hour of computing time. AWS sets a dynamically-changing ``spot price'' based on the current supply and demand.\footnote{AWS provides an API to provision both individual machines and in bulk (``spot fleet''). At the time of our demonstration, our underlying provisioning tools did not support spot fleet.} If the user's bid price is above the spot price, and there is sufficient capacity in the resource pool, the resources are provisioned at the spot price. If the spot price fluctuates above the bid price after a resource has been provisioned, the user is preempted with a two minute advance notice. Resources are charged on the hour boundary and when the instance is terminated by the user; in the event of preemption, the last fraction of an hour is not charged.

\subsubsection{AWS Limits}
In order to ensure that the AWS service is properly scaled to support the user workflow and to control potential runaway costs, the service has a number of adjustable global default limits, some of which are hidden to the end-user. We encountered several of these during testing and the initial ramp-up of the CMS workflows.

The Elastic Network Interface (ENI) service manages the network interfaces on the AWS worker nodes. A limit was discovered when the number of instances exceeded the default allocation and new Squid/Frontier instances began to generate errors when attempting to provision additional ENI. The default limit is dynamically generated, but can be statically set upon request; the limit was increased to 5000.

The HEPCloud team was notified ahead of time by AWS that the use case would require an increase in Elastic Block Storage (EBS) limits. We initially expected to use an Amazon Machine Image (AMI) with a single 7 GB EBS volume for the operating system, and two ephemeral disk volumes. After gaining some experience with the AWS service, EBS-only AMIs were added to the list of provisionable resources, as some instance types
do not support ephemeral volumes. Increasing the diversity of instance types reduced the overall preemption rate and took advantage of good price-performance. This required an additional EBS limit increase request from 20 TB to 300 TB per region.

The limits governing the number of spot instance requests per region had to be significantly increased over the defaults---from 20 to 5500---in order to scale to 60,000 cores. As a precaution, the limits governing the number of non-spot instances were lowered to 20, the expected bounds of the on-demand Squid/Frontier instances.

Additionally, we had to raise the limit on the number of entries per Security Group from 50 to 125. The project, in fact, established a requirement for a ``deny all'' security posture in order to reduce the risk of errant jobs running up outbound network costs, as described in the previous section.
Since we were accessing S3 over its public network interface, we had to explicitly enable outbound access from our instances to the S3 endpoints in each region. This access was granted by configuring a large number of whitelisted subnets in the Security Groups.

\subsection{Services Deployed at Fermilab}
\subsubsection{Accounting and Billing}
Fermilab operates an accounting system for the utilization of grid resources. We extended this service by developing an additional probe to collect usage data from Amazon. This probe polls the AWS monitoring interface every hour to detect the number of machines instantiated by instance type, the associated virtual organization and AWS account, and the spot price charged for that hour. For instances that have been terminated, it records the termination reason and time. This information is recorded in the Gratia \cite{gratia} database and the Gratiaweb service can be used to analyze and display it.

DLT Solutions supplied a billing summary of usage in comma-separated-value format and provided an hourly report on all AWS service usage. Custom routines were written to parse this billing data and keep track of our
balance, so that we could know how much remaining funds were available.  We also set up our own alarms to alert if the burn rate was unexpectedly high. Estimates of the data egress cost and its ratio to the total cost were also calculated. Because monthly data egress charges below 15\% of the total cost are not charged, this estimate informs operations of the potential costs of data egress. An access-restricted Grafana\footnote{The open platform for analytics and monitoring, https://grafana.com} instance was deployed for monitoring financial data.

\subsubsection{glideinWMS}
The glideinWMS workload management system \cite{Sfiligoi,Mhashilkar} was used to provision the worker nodes used during the CMS run. GlideinWMS
follows the pilot-based workload paradigm---launching a pilot to provision and validate a compute resource, which then pulls work
from a central queue.
A development version of glideinWMS was deployed to make available some of the new features needed to run at scale at AWS.  During the testing period prior to ramp-up, several patches were applied \textit{in situ} to address various issues found. These patches were provided to the glideinWMS development team for addition to later development releases.

The glideinWMS HEPCloud instance was configured with 120 resource types. Each type consisted of
AWS region in the US, availability zone, and instance type; additional parameters (such as the maximum instance lifetime) were added to the configuration.
The AMI ID and credentials needed to launch the instances are passed securely between glideinWMS components.

The instances were configured to run a bootstrap service. This service parsed a base-64 encoded string (``user-data'') that was passed to the instance when provisioned. The user-data contained the instance lifecycle parameters, the X.509 proxy used for daemon communication, and the glideinWMS pilot arguments (including the URL of the glideinWMS pilot scripts). The bootstrap service downloaded the pilot scripts and associated files, then launched the pilot within the instance.

The lifecycle of the provisioned instance had the following stages. The factory requested instances as prompted by the frontend. The maximum lifetime was passed to the instance as part of user-data and enforced by the system. The pilot exited when no more jobs match the resources or when the maximum lifetime was exceeded. The instance was configured to shut down when the pilot exited. Additionally, an administrator could issue commands on the factory to remove provisioned instances.

\subsubsection{HTCondor / Submission Pool}
At the time of the workflow testing and execution, the CMS global HTCondor pool \cite{htcondor,cms_condor} was not capable of increasing in scale by an additional 60,000 jobs. To address this, we bypassed the CMS global HTCondor pool and provisioned a separate HEPCloud HTCondor pool. We deployed three independent machines at Fermilab as schedulers and two additional nodes to serve as a highly-available HTCondor central manager. Instances of the CMS-specific workload submission system WMAgent \cite{wmagent} were deployed on each scheduler. The WMAgent instances retrieved descriptions of work from
a central service, created jobs, submitted them to their local batch HTCondor scheduler, tracked their completion, and resubmitted failed jobs when needed.

\subsection{Reporting}
The CERN Dashboard\footnote{http://dashboard.cern.ch} was used to track all the metrics collected by the workflow. Included in the workflow data were total runtimes for each step, data I/O, and efficiencies.

For infrastructure reporting and monitoring, Fermilab deployed an additional publicly available Grafana instance with customized views.
The views included aggregate numbers for AMI types per availability zone and region, number of cores provisioned, and a running count of instances in running, idle, and preempted state.

We also leveraged an already deployed ELK (ElasticSearch, LogStash, Kibana) stack\footnote{https://www.elastic.io} to read in all the data available from the local HTCondor job scheduler, allowing rich data mining of the HTCondor job data. Additional data were available via logfiles that the jobs stage out as part of their routine workflow.

\section{Costs and stability of services}

Motivated by a previous study \cite{Wu}, we selected a simple bid strategy for spot pricing, which was to bid 25\% of the ``on-demand'' price for a given resource.

\begin{figure}
\centering
 \includegraphics[width=.5\textwidth]{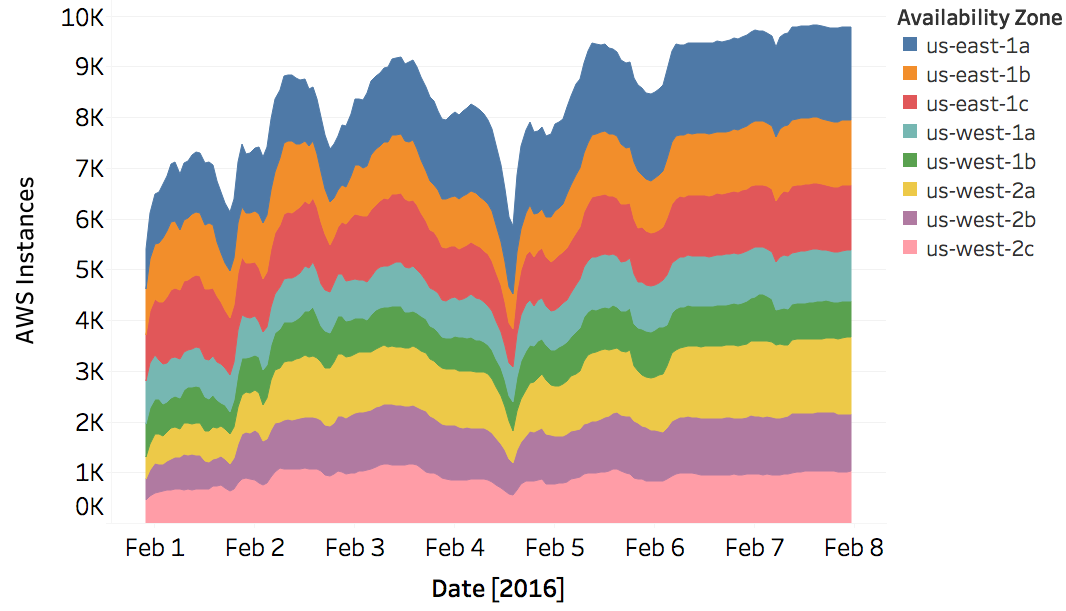}
\caption{\label{fig:zones}Number of running instances by AWS Region and Availability Zone.}
\end{figure}

In this study, various adaptive algorithms and strategies previously described in the literature were evaluated. The cost of each strategy was
computed, based on collection of the spot price history over an interval of four months.  The static strategy gave the most balanced performance in terms of cost
when averaged over a variety of pricing conditions.

We devised a strategy of ``portfolio diversification'' based on our observations of the spot market. To improve the availability and
stability of the system at scale, we bid in more than 100 different spot markets, representing nearly all the regions and zones then available in the US, as illustrated in Figure~\ref{fig:zones}.

\begin{figure}
 \centering
 \includegraphics[width=.23\textwidth]{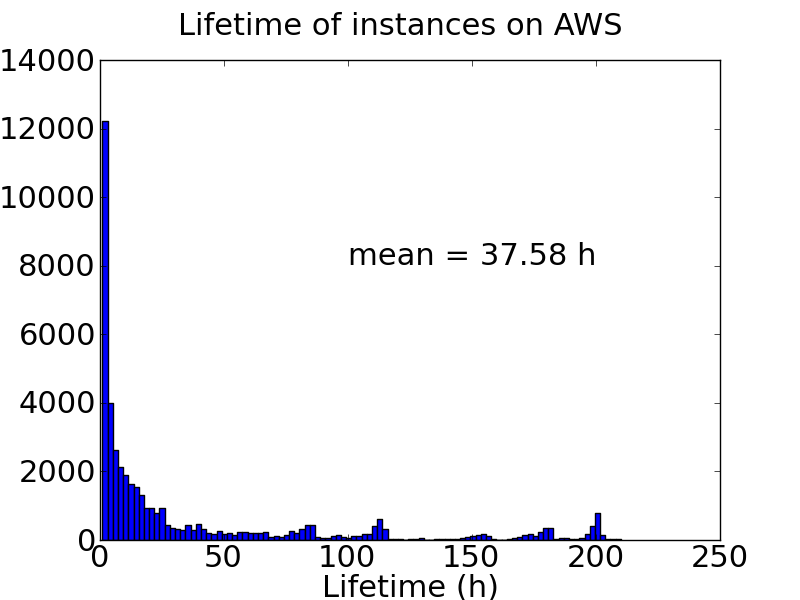}
 \includegraphics[width=.23\textwidth]{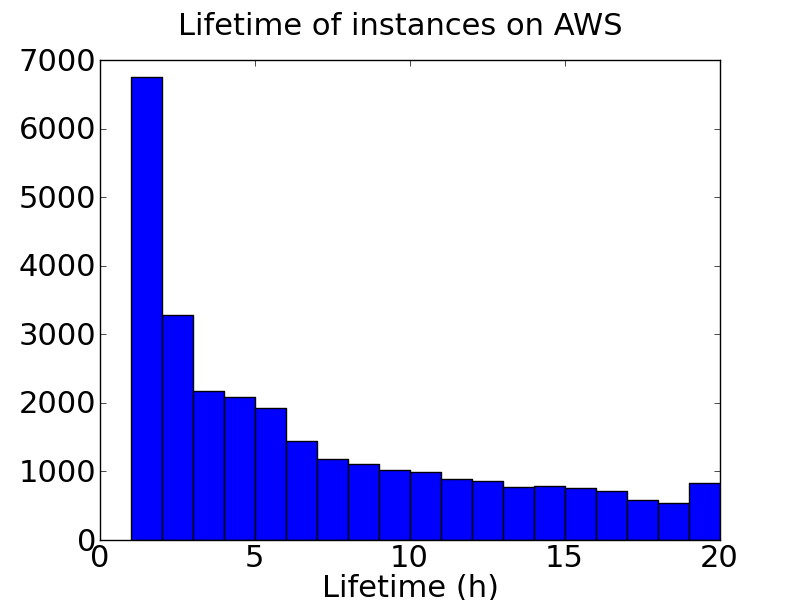}
\caption{\label{fig:pilot_lifetime}Pilot lifetimes in hours. The left histogram represents the entire distribution; the right histogram is zoomed in to show the distribution for pilots with lifetimes under 20 hours.}
\end{figure}

The mean lifetime for a provisioned resource was 37.6 hours, while the average job lifetime was 4.7 hours. Figure~\ref{fig:pilot_lifetime} shows the distribution of provisioned resource lifetimes. While the distribution is peaked in the lowest bin, the tail is very long---some resources remained in the pool for over 200 hours.

Over the course of the 3.2 million job run, 15.5\% of the jobs were preempted, as shown in Table~\ref{aws:preemption}. Preemptions are made visible within HTCondor as the disappearance of a provisioned resource.
When a preemption is detected, the scheduler reschedules the job and restarts it on a different available resource. The ``number of job starts'' is then strictly one less than the number of times a job was preempted.

\begin{table}
\caption{Preemption counts for CMS jobs}
\label{aws:preemption}

\begin{tabular*}{.5\textwidth}{@{}l*{15}{@{\extracolsep{0pt plus 12pt}}l}}
\\
Number of times preempted &	Count & Percentage of total \\
\hline
0 &	2736240 & 84.5\% \\
1 &	403062 & 12.4\% \\
$>1$ & 101687 & 3.1\% \\
\end{tabular*}
\end{table}

\begin{figure}
\centering
 \includegraphics[width=.5\textwidth]{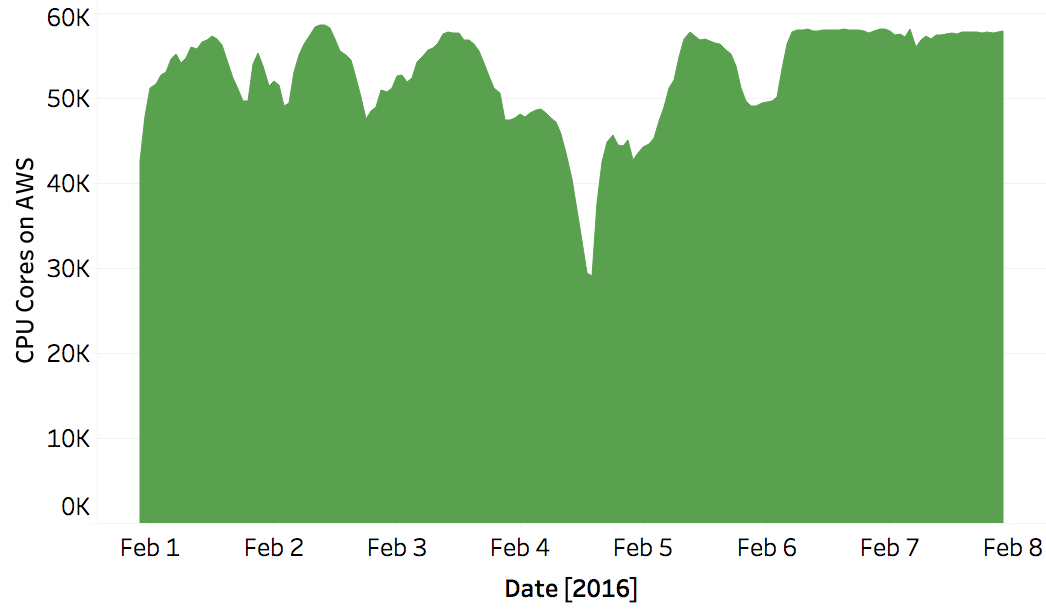}
\caption{\label{fig:pushing_scale}Count of CPU cores on AWS from February 1\textsuperscript{st} (Monday) to February 7\textsuperscript{th} (Sunday), 2016.
The plateau near 60,000 cores is limited by the local submission infrastructure. With the exception of the large dip on February 4, the decreases are purely due to the dynamics of the spot market.}
\end{figure}

There is also a time-dependence to the ability of acquiring resources at scale. In general, during the business day, AWS removes resources from the spot market to fulfill their ``reserved'' and ``on-demand'' classes of service. In the late evenings and on the weekends, as the demands on those classes of service go down, the supply of resources into the spot market increases. This is clearly visible in Figure \ref{fig:pushing_scale}---there are peaks in the early morning hours and on weekends, and valleys during the day when resources were removed from pools and machines were being more frequently preempted.

The default strategy used by the glideinWMS frontend and factory was to attempt to distribute the load evenly (on a number of core basis) across nearly all regions, zones and instance types\footnote{https://aws.amazon.com/ec2/instance-types/}. We found that some region/zone/instance type combinations filled up very quickly and the price quickly moved above our bid price, causing quick preemption of those instance types.  Over time we ended up accumulating most of the instances which were least likely to get preempted.
The final instance mix near the end of the steady-state is shown in Table~\ref{aws:instancetypemix}\footnote{Instance types that provided smaller contributions are not included.}.

\begin{table*}
\caption{AWS instance distribution at steady-state. (Actual pricing was at or below the bid price.)}
\label{aws:instancetypemix}
\begin{tabular*}{\textwidth}{@{}l*{15}{@{\extracolsep{0pt plus 12pt}}l}}
\\
Instance Type 	& Number of Instances	& Number of Cores per Instance   & Memory per Instance (GiB) & Bid price per Instance\\
\hline
m3.xlarge & 2905 & 4 & 15 & \$0.0665 \\
m3.2xlarge & 1244 & 8 & 30 & \$0.133 \\
r3.2xlarge & 1109 & 8 & 61 & \$0.175 \\
r3.xlarge & 826 & 4 & 30.5 & \$0.0875 \\
c3.xlarge & 759 & 4 & 7.5 & \$0.0525 \\
c3.2xlarge & 614 & 8 & 15 & \$0.105 \\
m4.xlarge &  655 & 4 & 16 & \$0.063 \\
m4.2xlarge &  413 & 8 & 32 & \$0.126\\
\\
\end{tabular*}
\end{table*}

During the run, \$211,985 was spent on AWS services. 15,085,635 wallclock hours were consumed---giving an average cost per wallclock hour of 1.4 cents. Approximately 92\% of cost was spent on EC2 instances, 6\% on support, and 2\% on S3 storage.  The cost per event for different physics samples is shown in Table~\ref{aws:costperevent}.

\begin{table*}
\caption{Average time and cost per job and event, by sample}
\label{aws:costperevent}
\begin{tabular*}{\textwidth}{@{}l*{15}{@{\extracolsep{0pt plus 12pt}}l}}
\\
Sample 	& Average time per successful job (s)	& Average time per event (s) &	Cost per 100 events \\
\hline
TTJets     &    25,345 & 42.2 & \$0.016 \\
DY\_M10--50 &	14,111 & 23.5 &	\$0.0092 \\
DY\_M50    &	13,214 & 22.0 &	\$0.0085 \\
WJetsToLNu &    12,235 & 20.4 &	\$0.0079 \\
\\
\end{tabular*}
\end{table*}

\section{Lessons Learned and Operational Considerations}

At the beginning of the CMS AWS HEPCloud use case, a significant effort went into looking at the monitoring that would be needed to prevent unnecessary waste of computing resources. Tracking slow and stuck jobs, identifying infinite loops in the application, identifying I/O-bound jobs and other sources of low CPU efficiencies, and protecting against huge log files that would incur high export transfer charges, were all considered. On AWS, the financial loss associated with inefficiency is explicit, but an early conclusion of the AWS investigation was that we should ensure the monitoring put in
place for elastic resources is extended to dedicated grid resources.
On dedicated computing that has been purchased in advance, it is easier to mislead yourself that occurring  inefficiencies are not a financial loss,
but the issues are the same as with cloud resources; the costs have just been amortized up-front and are hidden.

AWS imposes a substantial fee for data egress out of the cloud.
Fees begin at \$0.09 per GB and drop to \$0.07 per GB as the total egress per month exceeds 100 TB (at the time the use case was executed).
As mentioned earlier, these charges are waived if the egress charges are less than 15\% of the total processing charges.
Optimizing the use of resources encourages longer running workflows with small output; there are no charges for import.
By chaining together {\it GEN-SIM} and {\it DIGI-RECO} into a {\it GEN-SIM-DIGI-RECO} workflow, we kept egress charges under the 15\% threshold for the use case. Specifically, each job staged back approximately 161 events, and the average event size was about 216 kB.

The I/O characteristics of the experimental workflow impacted the operational costs. In the {\it DIGI} stage of the workflow,
the job executed a loop that performed some computation, did hundreds of reads from input data, and then repeated. The initial strategy was to use S3 storage for the input data, and to do streaming reads from the
application.
However, there is a small charge for each read (seek) operation. We discovered that the CMS application performed so many read operations (150 million HTTP GET requests per hour) that the cost of I/O was
comparable to the costs of processing time. As a result, we changed strategies to retrieve an entire input data file to the local worker node storage
and read it from there. The change reduced the cost by five orders of magnitude and increased the job efficiency.

As discussed above, the system was configured to provision multiple resource types from multiple Amazon regions and availability zones to increase the total scale of available resources on the spot market. Data, however, was stored at a specific Amazon region. To access the data, therefore, one could opt to store the dataset in one region, for about \$0.03 per GB per month, and access it from all regions, incurring in inter-region transfer charges at \$0.02 per GB. Alternatively, one could replicate the dataset in all regions, increasing the storage charges, to avoid inter-region transfer charges. The latter was the most cost effective strategy for the size of the input data.

We also needed to ensure that only authenticated users could read from S3, as we could not control the quantity of data being read (and charged) for unauthenticated access. Read privileges were granted to the compute nodes via ``roles'' which were assigned at the time
of node instantiation. The compute node then retrieves private, public, and session keys via the AWS Security Token Service (STS); this method is preferable to the less-secure method of passing in private keys from outside AWS with appropriate privileges. In order to stage in the complete file to the node, we leveraged the existing callout to a curl executable in the CMS software framework \cite{CMSSW}.
Two additional binaries were deployed to the worker node via the glideinWMS frontend---an AWS-capable curl that generates custom AWS HTTP authentication headers, and another that wraps the
environment configuration in order to prepend an appropriate path to the environment. In addition, a bootstrap script tied into the standard CVMFS setup commands was added to the local
image.\footnote{https://github.com/holzman/glidein-scripts}

\section{Performance and reliability}

Table~\ref{aws:jobfailures} shows the job failure rate as seen by the WMAgent system for the various samples produced in {\it GEN-SIM-DIGI-RECO} workflows.
These rates do not account for intermediate failures that are retried by the underlying HTCondor batch system (e.g.\ due to node preemption), or the built-in 3-try resubmission from WMAgent itself.
Except for the DY\_M50 outlier (where the reason is understood, and the events were eventually recovered), the job failure rate looks competitive or even better than what one would expect if the workflow was running at CMS-owned sites.

\begin{table}
\caption{The number of events, jobs, and failure rates of each of the 4 large-scale {\it GEN-SIM-DIGI-RECO} workflows. The outlying failure rate (labeled with *) was
due to overloading the Fermilab local storage.}
\label{aws:jobfailures}
\begin{tabular*}{.5\textwidth}{@{}l*{15}{@{\extracolsep{0pt plus 12pt}}l}}
\\
Sample & Events & Jobs & Failure Rate \\
\hline
DY\_M10--50 & 121 M & 560590 & 0.6\% \\
DY\_M50 & 	65 M   & 185971 & 14\%* \\
TTJets &   197 M   &1089459 & 0.5\% \\
WJetstoLNu & 66 M  & 459184 & 0.1\% \\
\hline
\end{tabular*}
\end{table}

\begin{figure}
\centering
 \includegraphics[width=.5\textwidth]{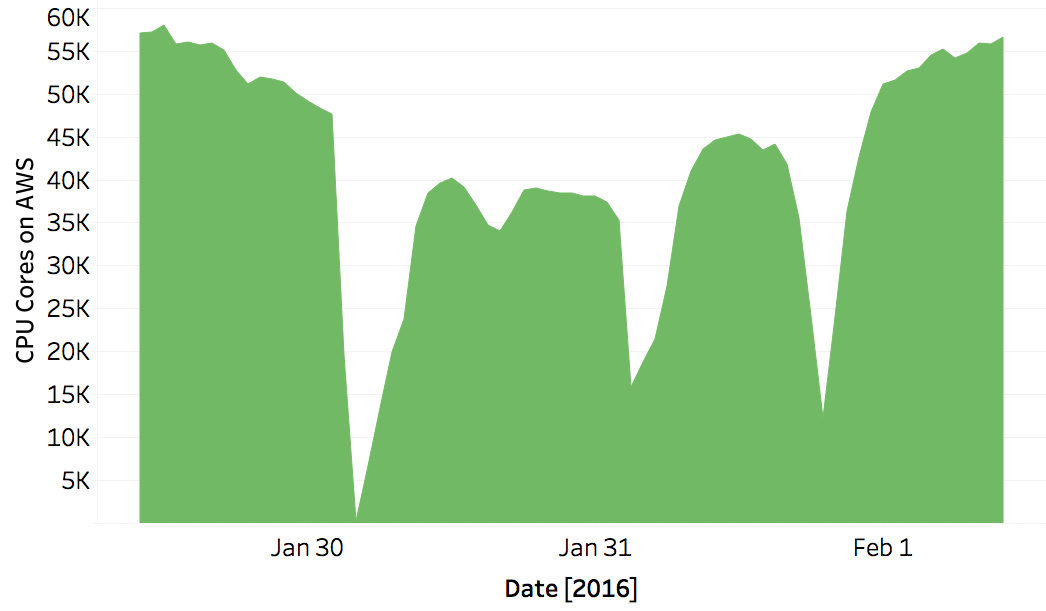}
\caption{\label{fig:stable_working_point}Count of CPU cores on AWS from January 28 to February 1, 2016. Valleys in the distribution
correspond mostly to failures on components external to AWS.}
\end{figure}

The reasons jobs can fail are many, but assuming the workflow itself is stable, the failures are usually related to external dependencies. Among those,
the most common are I/O related failures---reading input or staging output to mass storage. Figure~\ref{fig:stable_working_point} shows the effect of some of these failures on the job numbers during the time when we were trying to find a stable working point.

We encountered several issues. The Fermilab storage---destination for the output data---was overloaded at full scale. The implementation of the transfer protocol did not scale well in the storage system; we remedied this by switching to a better-supported protocol.\footnote{The EOS storage system \cite{eos} implements the SRM protocol by deploying the BeStMan \cite{bestman} software package, which is not well-supported. We switched to using the xrootd protocol, which
is supported natively by EOS.}
An image was misconfigured, causing failed reads of input data from S3; a properly-configured image was rolled out.
The security token used to authenticate reads was expiring too soon; this was corrected by acquiring the token ``just-in-time'' rather than when the instance was
provisioned.

Apart from failure-related drops in job count, both Figure~\ref{fig:pushing_scale} and Figure~\ref{fig:stable_working_point} show that it is not trivial to keep 50,000 cores fed with jobs continuously. We had two occasions where problems with the WMAgent service prevented us from keeping the AWS resources busy: we couldn't keep up with the required job submission rate.
Because glideinWMS only provisions resources when there are idle jobs in the queue, and because we also terminate provisioned resources after a short
time if they are idle, this did not incur a large cost inefficiency. In short, these issues did not cause any failures \textit{per se}, but had an impact on how quickly we could finish a workflow running on AWS.

Another source of inefficiency in the system is due to the spot market. Our jobs could get preempted if the cost of the resource exceeds the maximum bid that CMS is willing to pay for the resource. This would not show up as a failure, since HTCondor will reschedule the job. However, it has a similar effect as failed jobs---it causes computing resources to be spent on computations for which we get no usable output. Table~\ref{aws:times} shows the sum of wall clock and CPU time for jobs run between January 13 to February 12. The first number represents all jobs, including ones that are preempted and retried. The second and third are sums for only the final iteration of a job that ran to completion (including failed jobs).

\begin{table}
\caption{Wall clock and CPU time totals for CMS jobs on AWS}
\label{aws:times}
\begin{tabular*}{.5\textwidth}{@{}l*{15}{@{\extracolsep{0pt plus 12pt}}l}}
\\
Sum of all jobs Wall Clock (h) & 15,087,067 \\
Sum of completed jobs Wall Clock (h) & 13,663,074 \\
Sum of completed jobs CPU Time (h) & 11,885,993 \\
\hline
\end{tabular*}
\end{table}
From these numbers we can deduce that there was a roughly 10\% inefficiency due to preemption losses (but the monetary losses are lower because of the way AWS bills in the case of preemption). The other deduction is that average CPU efficiency over all final job iterations is 87\%. This is a very good number considering that our workflows run every step of {\it GEN-SIM-DIGI-RECO} sequentially on the instance and not all these steps are CPU-bound. For reference, the average efficiency for {\it GEN-SIM}, {\it DIGI}, and {\it RECO} on the grid are 57\%, 68\%, and 82\%, respectively. % \cite{sciaba}.

\section{Detailed cost comparison}

Cost is one of the most interesting comparisons between commercially provisioned resources and dedicated purchased computers. Historically, commercially provisioned computing has been much more expensive than regularly used purchased systems.  In recent years, due to the market competition, there has been a steady decrease in the cost of commercial computing. This decrease, combined with the evolution of spot pricing as a feasible working model, has made the commercial sources of computing more cost competitive.

Fermilab attempted to estimate the cost per core-hour of the CMS Tier-1 processing resources.
In this calculation there is a number of objective (but site-specific) values, such as the cost per kilowatt of power, the initial cost of the machines, the average lifetime for computers, and the amortized cost of the computing center building. There are also several more subjective inputs, such as the effort required to perform the administrative functions and the average utilization of the dedicated systems. In the cost calculations it was assumed that 3 Full Time Equivalent (FTE) units of effort were required to handle the local network and hardware administration of approximately 700 computing systems.
The estimate assumes 100\% utilization of CMS  Tier-1 resources; at lower utilization, the effective cost per productive CPU cycle is larger.
Given the uncertainty in the subjective inputs, we estimate that the error on the per hour core cost is roughly 25\%. There are also several assumptions in this
estimate. As a national laboratory and host to an accelerator complex, Fermilab buys electrical power in bulk at favorable prices. The costs also do not
include the price of constructing new data centers.

\begin{table*}
\caption{The cost per hour for one core of computing on dedicated Tier-1 resources at Fermilab and on virtualized commercial cloud resources on AWS and the $t\overline{t}$ benchmark (greater = faster). The uncertainty in the AWS cost data corresponds to one standard deviation from the daily cost per core-hour. The AWS cost does not include cost of staff.}
\label{aws:cost}
\begin{tabular*}{\textwidth}{@{}l*{15}{@{\extracolsep{0pt plus 12pt}}l}}
\\
Site &	Average cost per core-hour & $t\overline{t}$ benchmark ($t\overline{t}$ / s per core) \\ \hline
Fermilab CMS Tier-1 & \$$0.009 \pm 25\%$ \cite{Fuess} & 0.0163\\
AWS & \$$0.014 \pm 12\%$  & 0.0158\\
\hline
\end{tabular*}
\end{table*}

An important consideration when comparing the relative costs of core-hours is ensuring that the work performed by each core in an hour is also comparable. CMS performed a series of benchmarks using a standard simulation workflow---the so-called $t\overline{t}$ benchmark---comparing the speed of event production on a variety of AWS instances and several generations of hardware at Fermilab \cite{chep2016_timm}. Looking at the same number of cores used by the application, there is a spread of roughly 30\%. Newer hardware generally tends to be faster for event production even if it has a lower clock speed. The spread of performance is observed at both locations and there is virtually no systematic difference between the benchmark performance on local Fermilab equipment versus AWS-hosted systems. The performance is similar, so we believe it is reasonable to directly compare the core costs.

As shown in Table~\ref{aws:cost}, commercially provisioned resources are roughly 50\% more expensive than dedicated well-utilized local resources, but there are some caveats.   Not all workflows could be performed on AWS at this low cost rate.   Workflows that have large output and incur high export charges would be more expensive, and workflows that require large random access to data not available within AWS would be less efficient and therefore more expensive. A similar test at a lower scale using the NOvA experiment \cite{nova} measured an average cost of \$0.03 per core-hour because they needed to use larger systems and transfer data between regions.  There is the potential for significant variation in cost.  On the other hand, Fermilab is only this inexpensive if the resources are continuously used.   One of the most attractive elements of commercially provisioned resources is that they can be dynamically provisioned.   There is no difference in cost to require twice as much computing in one month and nothing in the following month. The advantages of this kind of peak scheduling will be discussed in the following section.

Given the continued evolution to lower costs in commercially provisioned computing resources and the success of this project utilizing them, it is likely that the HEP community will pursue a hybrid utilization model with contributions from both dedicated grid computing and commercially provisioned computing. There are differences in the workflows that are best suited for each, so it is unlikely that one will completely replace the other, but a model with complementary resources is both desirable and cost effective.

\section{Non-economic value of commercially provisioned resources}

\begin{figure}
\centering
 \includegraphics[width=.5\textwidth]{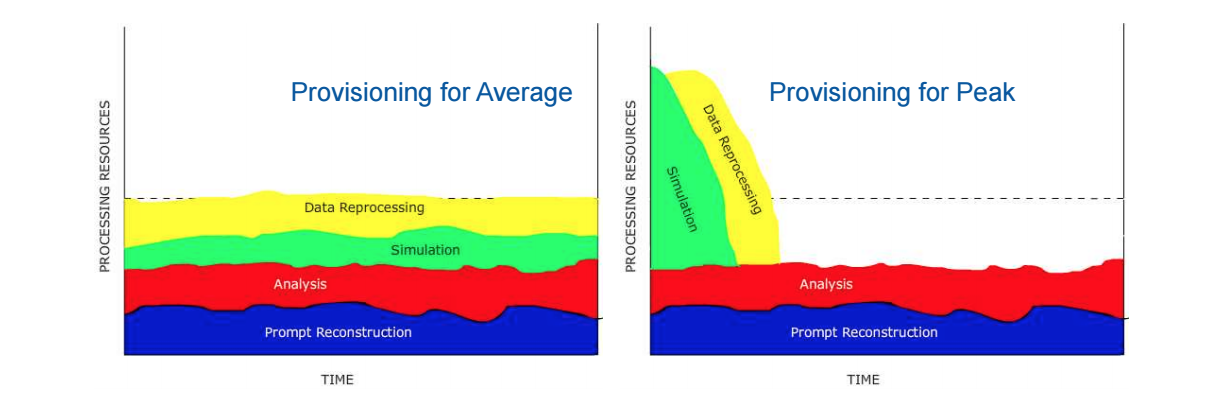}
\caption{\label{fig:peak}An illustration of provisioning for average vs. provisioning for peak.}
\end{figure}

The computing resources for HEP experiments are pledged by the funding agencies of countries participating in the Worldwide LHC Computing Grid (WLCG \cite{WLCG}).
Because of the lead time to commission physical resources in the computing centers in the different countries, the planning process looks 18 months ahead.
The HEP experiments plan and request computing resources yearly. Requests are scrutinized and eventually endorsed in a formal process by the Computing Resource Scrutiny Group (C-RSG).
Resources are deployed on a specific date, and then available to the experiment throughout the year; it is very important that the experiments'
central production teams plan for steady and continuous use during long periods of time,
as shown on the left side of Figure~\ref{fig:peak}. Experience from Run1 and Run2 at the LHC shows that the computing needs of experiments are not constant over time.
A number of activities, such as data (re)processing, simulation data generation and reconstruction, tend to come in bursts with irregular time structure, dictated by software release, conference and data taking schedules. Thus, there is a significant mismatch of the planning frequency, and thus, provisioning of resources, and the frequency with which user demand varies. In the example of a conference deadline, production activities have to start well in advance to make the deadline with constant resource usage.   Incorporating elastic resource provisioning (e.g.\ commercial cloud) could enable a much more efficient processing plan starting shortly before the conference. The available elasticity of bursting resources into commercial clouds would change the way people work in large scientific collaborations and allow for shorter and more agile time schedules. The right side of Figure~\ref{fig:peak} shows this case where processing and simulation is done in burst. With resources provisioned with commercial clouds, the planning process could also be condensed. Time to provision resources is shorter because physical resources do not have to be provisioned and installed at the computing centers.

Provisioned resources like AWS may also provide a powerful source for problem recovery. In case of a problem that invalidates work already performed (either due to software, a systematic computing issue, or not properly accounting for changing experiment conditions) there is not sufficient excess capacity in the system to perform the work twice, without having to make difficult choices to cancel needed future work. At the same time it is not possible in the current budget environment to reserve excess capacity to recover from this type of failure. The cloud model is interesting because it allows for the dynamic purchase of sufficient capacity to solve problems without maintaining dedicated resources in reserve. This ability to burst to a high fraction of the total CMS resources for a period of time should be seen as an useful insurance policy to recover from such  problems.

Specialized resources like high-memory slots and other more exotic hardware configurations might be provisioned more flexibly with commercial partners. This would allow to react flexibly and maximize physics output without long-term investment in physical hardware.

\section{Looking Forward}

\begin{figure}
\centering
 \includegraphics[width=.5\textwidth]{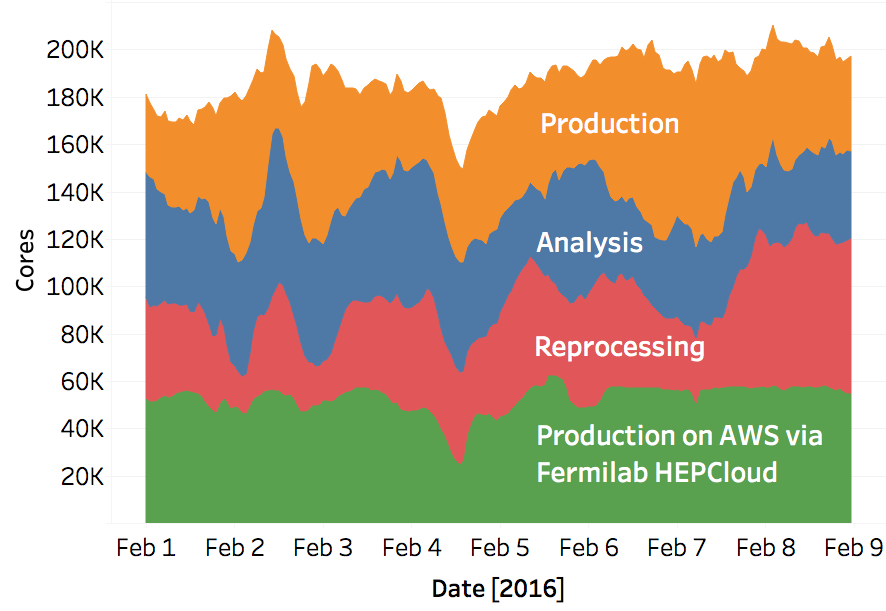}
\caption{\label{fig:aws_cms_global}A comparison of the scale of processing on AWS to other global CMS activity.}
\end{figure}

\begin{figure}
\centering
 \includegraphics[width=.5\textwidth]{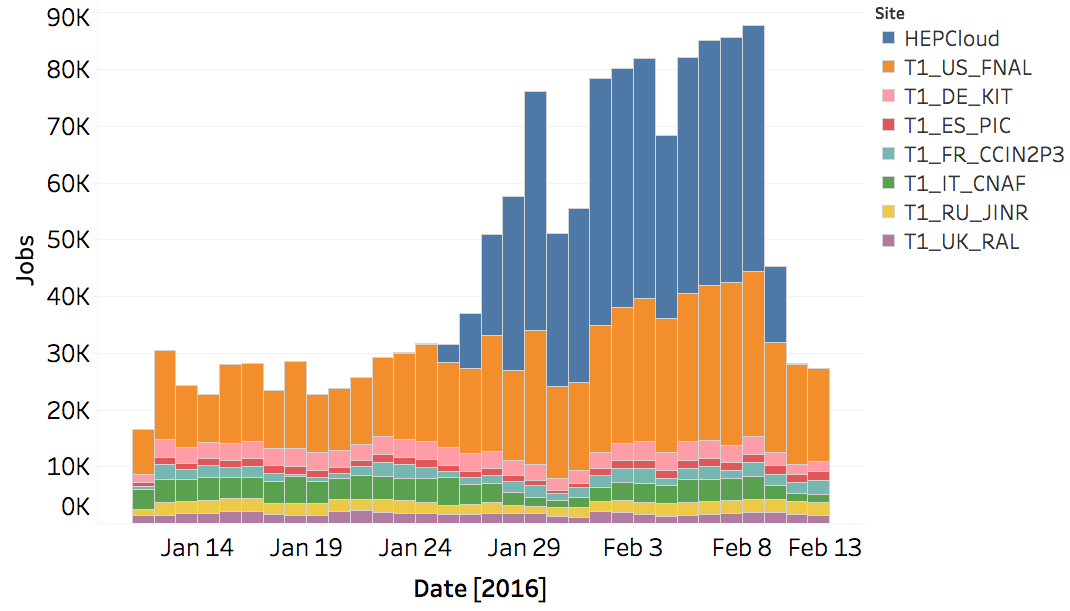}
\caption{\label{fig:aws_t1}A comparison of the scale of processing on AWS to other CMS Tier-1 activity.}
\end{figure}

The HEPCloud CMS use case on AWS, performed by Fermilab and CMS, has demonstrated that it is possible to utilize dynamically provisioned cloud resources to sustain execution of many CMS workflows at an extremely large scale. As shown in Figure~\ref{fig:aws_cms_global},
the HEPCloud facility was able to increase the amount of resources available to CMS by 33\%. When viewed in terms of the
expansion of the Tier-1 facilities, as shown in Figure~\ref{fig:aws_t1}, the effect is even larger.

In order to operate the resources most efficiently we selected specific workflows, but we did not find any that could not be executed.   Individual sites that consider purchased computing services as a component of their processing pledge to the experiment should be able to maintain a production efficiency similar to what we observed, by deploying dedicated interface systems for grid services on top of dynamically provisioned resources (e.g.\ through a HEPCloud-like solution).
\section{Global Context of Dynamic Resources}

The successful completion of real-life CMS workflows at scale as described in this note demonstrates the potential for utilizing the HEPCloud paradigm and establishes the merits of this approach. Commercial cloud resources rented from AWS (supported through an AWS grant) were successfully integrated into the current Fermilab computing facility in a manner transparent to the experiment.

The CMS HEPCloud use case with AWS demonstrated scalability at a level of the worldwide LHC computing scale. This result is crucially important to  extrapolate to the expected future exponential increases in computing needs, establishing that cloud provisioning is becoming a real contender for realizing these future needs.

We demonstrated that elasticity of cloud provisioning is very high, certainly sufficient to address the bursting needs of HEP computing. We also demonstrated that this approach is compatible with the operational procedures of a number of diverse experiment workflows. This approach can be used in ``production mode'' once the HEPCloud portal and the flexible services backbone will be fully functional and running in operations mode.  HEPCloud will then provide important new on-demand capabilities to experiment production managers; these capabilities will increase flexibility and enable efficiency gains in the experiment resource planning process.

We demonstrated that performance and efficiencies are high, comparable to and sometimes surpassing dedicated HEP resources. We measured CPU efficiency to be approaching 90\% overall, which compares favorably with dedicated resources even given the rather complex workflows used in the demonstrator. The demonstrated reliability and high availability of the ``as-a-service'' approach is sufficient to serve as a backup and insurance against eventual local outages, potentially increasing the overall robustness of the local facility.

Cost effectiveness is a complex issue, but given the increasingly competitive market of cloud providers we expect it to further increase. HEP can make use of spot market prices effectively. We saw that preemption caused only a 10\% inefficiency overall, and the cost impact of this inefficiency was actually much lower, given the AWS spot market pricing policies. The cost comparisons assume almost 100\% utilization of owned resources, a value that is rarely reached or sustained over the year, given the inability to plan at the frequency required for full resource utilization, the ``burst'' nature of experiment computing workloads, and the inherent ``inelastic'' nature of owned resources. As a matter of fact, the LHC utilization history of Tier-1 resources has been significantly lower, narrowing the cost gap between on-premises and off-premises further.

The current AWS costing model puts a premium on data transfers, making data intensive workflows a cost driver. However, this may change in the future, given that the actual networking cost per unit data continues to decrease exponentially over time. HEP, via ESnet ``Points of Presence'' into the AWS cloud, has access to a flexible and high performance infrastructure of data access points, which should bring down the actual data transport costs to the provider. Using data transfer volume as a cost driver is part of the AWS business model today, but it is at least conceivable that future cost models for large-scale clients like HEP could de-emphasize data fees while still providing enough margin for cloud providers to be profitable.

The demonstrator described in this note is just the beginning of making HEPCloud a dependable part of the Fermilab and HEP computing infrastructure for the LHC, the intensity frontier and the neutrino program. For the LHC in particular, the new capabilities of on-demand resources and resource elasticity provided by HEPCloud are significant enough to partially outweigh the larger cost per core-hour, making HEPCloud provisioning of LHC resources a new and important ingredient to the overall computing ecosystem. HEPCloud adds commercial or community cloud resources to the predominance of owned resources, augments opportunistic resources across OSG, and contributes future HPC and supercomputing center resources to the mix. This approach will help HEP facilities move away from standalone, siloed solutions.

We expect the new capability of reliable and robust on-demand provisioning of HEPCloud to significantly impact future resource planning for the LHC and the rest of the HEP computing program. For the next cycle of computing resource planning, on-demand capabilities will lower the need for owned resources on the floor, lowering the need for over-provisioning as a strategy to deal with peak demands. The corresponding decrease in investment cost for owned resources will provide some flexibility for on-demand capacity. The exact balance for the coming resource years will need to be carefully determined. In the meantime, making the HEPCloud portal a robust and fully supported piece in the US HEP computing landscape will be of high priority and importance.

We  anticipate that the HEPCloud portal concept will provide a means for all laboratories to provide shared resources in the ecosystem, resulting in a large pool of offerings for compute, archival capabilities, database services, data management, etc., potentially linking all US HEP computing.
The concept, inception, and evolution is guided in large part by
the requirements of large international experiments (LHC, for example).
As such, it will be directly applicable and beneficial to the international
community, and could be extended to incorporate any additional requirements
introduced by the international aspect
(assuming international partners and funding).

\section{Conclusions}

The HEP experimental program continues to evolve, and will require computing capacity in excess of scaling current on-premise resources within reasonable budget scenarios and computing technology evolution. To help address this problem, it is essential that
we develop the capability to expand HEP facilities beyond what is on the local data center floor. A sensible target which will yield significant benefits is to leverage the industry trends in cloud computing. The HEPCloud facility concept provides a portal to such computing resources as a transparent
layer for the users, offloading the decisions of when and how to acquire off-premises resources to the facility and its Decision Engine.

We deployed an implementation of aspects of the HEPCloud facility concept using glideinWMS technology, with the goal of evaluating the performance and quantifying the benefits of the concept.  To achieve this goal, we designed and executed a number of different use cases for different experiments. The goal of the
CMS use case was to enable the execution of a physics workflow that would add significantly to their overall global compute capacity, while generating useful analysis results for the experiment.  This second aspect ensures that the use case demonstrates directly the benefits of the approach.
For a full simulation workflow (from event generation to physics reconstruction), over 15 million hours of computing were consumed,
simulating more than 500 million events. The steady-state cost came to $1.4 \pm 12\%$ cents per core-hour, which is not much larger than the estimated
$0.9 \pm 25\%$ cents per core-hour for the Fermilab data center. The NOvA experiment also executed a smaller-scale workflow on AWS (at a cost of 3 cents per core-hour),
demonstrating the ability of the HEPCloud facility to serve different user communities and exercising a different processing-to-data ratio.

From this work, we have shown that commercial cloud resources can be acquired at large scales for costs that are larger than, but comparable to, the cost of
procuring and deploying similar resources on-site. Given the large year-over-year increases in the size of cloud computing industry-wide and the potential
economies of scale, it is conceivable that the steady-state computing costs could approach or even undercut the price of procuring physical equipment. Beyond the comparison of steady-state costs, the needs and demands of the scientific community are not flat with respect to time, but have a structure and time-dependence.
Having the ability to pay for only the resources that are used gives a large amount of flexibility and can increase planning flexibility, efficiency,  and cost
effectiveness overall.
On the other hand, not all workflows are well-suited to running off-site. Despite the international proliferation of high-bandwidth networks, scientific workflows that are very data intensive may be better matched to executing on local resources near storage.  It is clear that a hybrid approach, the HEPCloud facility---capable of provisioning both on-premises and off-premises resources and aggregating them into a single virtual facility---will give the most flexibility and gives the
scientific community the best chance to meet the ever-growing needs of its users.

\begin{acknowledgements}
This work was partially supported by Fermilab operated by
Fermi Research Alliance, LLC under Contract No.  DE-AC02-07CH11359 with the United States
Department  of  Energy,  the  National  Science  Foundation  under  Grant  ACI-1450377,
Cooperative Agreement PHY-1120138, and the AWS Cloud Credits for Research program.
On behalf of all authors, the corresponding author states that there is no conflict of interest.
\end{acknowledgements}

\end{document}